\documentclass[aps,prl,twocolumn,showpacs,superscriptaddress]{revtex4}
\usepackage{amsmath}
\usepackage{amssymb}
\usepackage{color}
\usepackage{graphicx}
\usepackage{graphicx,amsmath}
\usepackage{bm}

\begin{document}
\newcommand{\beq}{\begin{equation}}
\newcommand{\eeq}{\end{equation}}

\title{Fate of the Wiedemann-Franz law near quantum critical
	points of electron systems in solids}
\author{V.~A. Khodel}
\affiliation{NRC Kurchatov Institute, Moscow, 123182, Russia}
\affiliation{McDonnell Center for the Space Sciences \& Department of Physics,
	Washington University, St.~Louis, MO 63130, USA}
\author{J.~W.~Clark}
\affiliation{McDonnell Center for the Space Sciences \& Department
	of Physics, Washington University, St.~Louis, MO 63130, USA}
\affiliation{Centro de Ci\^encias Matem\'aticas, Universidade de Madeira,	9000-390 Funchal, Madeira, Portugal}
\author{V.~R. Shaginyan}
\affiliation{Petersburg Nuclear Physics Institute, NRC Kurchatov Institute,
	Gatchina, 188300, Russia}
\affiliation{Clark Atlanta University, Atlanta, GA 30314, USA}
\author{M.~V. Zverev}
\affiliation{NRC Kurchatov Institute, Moscow, 123182, Russia}
\affiliation{Moscow Institute of Physics and Technology, Dolgoprudny, Moscow
	District 141700}

\begin{abstract}
	
We introduce and analyze two different scenarios for violation of the
Wiedemann-Franz law in strongly correlated electron systems of solids,
close to a topological quantum critical point (TQCP) where the density
of states $N(0)$ diverges.  The first, applicable to the Fermi-liquid
(FL) side of the TQCP, involves a transverse zero-sound collective mode
that opens a new channel for the thermal conductivity, thereby enhancing
the Lorenz number $L(0)$ relative to the value $L_0=\pi^2k^2_B/3e^2$
dictated by conventional FL theory. The second mechanism for violation
of the WF law, relevant to the non-Fermi-liquid (NFL) side of the TQCP,
involves the formation of a flat band and leads instead to a reduction of
the Lorenz number.

\end{abstract}

\pacs{71.10.Hf, 71.27.+a, 71.10.Ay}
\maketitle

\paragraph{Introduction.}
The perplexing issue of a possible violation  of the Wiedemann-Franz (WF)
law in strongly correlated electron systems of solids has recently
attracted much attention. The WF law states that as the temperature
$T$ goes to zero, the product $L(T)=\kappa\rho/T$ of the thermal
conductivity $\kappa$ and the electrical resistivity $\rho$, divided
by $T$, always approaches the Lorenz number $L_0=\pi^2k^2_B/3e^2$.
Over a brief period, several articles have appeared
\cite{steg,flouq,tai,arxiv,brison,reid}
whose authors reported or discussed putative evidence of a breakdown
of this fundamental law in the antiferromagnetic (AF) heavy-fermion metal
YbRh$_2$Si$_2$. This compound has become one of the prime test cases in
the quest for understanding of NFL behavior of electron systems that
exhibit quantum critical points (QCPs) at which lines of second-order
phase transitions are terminated.  Conclusions drawn about the fate of
the WF law have varied, but the authors of Ref.~\cite{steg} assert
that the WF law is definitely violated.

When one considers the possibility of violation of the WF law, it must
be emphasized that this law is, in fact, well satisfied across the
wide range of metallic electron systems that have been studied
experimentally, except for two particularly interesting compounds
in addition to YbRh$_2$Si$_2$.  The first is the normal state of the
superconducting heavy-fermion metal CeCoIn$_5$ \cite{tanatar}
and the second, the heavy-fermion compound YbAgGe \cite{ dong}.
QCPs of the type indicated above are present in the phase diagrams
of all three of these electron systems.

It is important to recognize that in all the cited publications
claiming that the WF law is definitely or likely to be violated,
the temperatures reached in the measurements involved were no lower
than $40\, {\rm mK}$.  However, in recent studies of YbRh$_2$Si$_2$,
measurements have been performed at considerably lower temperatures
$T\simeq 10\,{\rm mK}$.  Most significantly, the thermal resistivity
$w(T)=L_0T/\kappa(T)$ was observed to execute an unexpected downturn
at $T\leq 30\, {\rm mK}$, shown in Fig.\,1 of Refs.~\cite{arxiv,brison}.
This behavior confounds any strong conclusions previously made about
the fate of the WF law in this compound and casts doubt on others.

\paragraph{Preliminaries: relevant phenomena in the quantum critical regime.}
According to the authors of Refs.~\cite{arxiv,brison}, the occurrence
of the downturn in the thermal resistivity $w(T)$ points to the
presence of an unknown collective mode, since all phonon contributions
have disappeared already by $T\leq 1\,K$.  One might then suppose
that in the AF state of YbRh$_2$Si$_2$, which terminates at
$T_N \simeq 70\, {\rm mK}$, magnons can provide the anticipated boson
mode.  This is not the case, however, since the downturn in $w(T)$
observed at $T\leq 30\, {\rm mK}$ persists in external magnetic
fields whose magnitude is in excess of a critical value sufficient
for termination of the AF ordering \cite{arxiv,brison}.

On the other hand, there is a special feature of the phase diagram
of YbRh$_2$Si$_2$ that is relevant to a search for the posited boson
mode.  Namely: the AF QCP resides close to a topological quantum
critical point (TQCP) associated with the divergence of the density
of states $N(0)$, which in a homogeneous Fermi liquid is proportional
to the effective mass $M^*$. The two critical points, i.e., the TQCP
and that of the termination of the AF state at $T_N$, {\it coincide}
in the presence of a tiny external magnetic field $B_c\simeq 70\,
{\rm mT}$ \cite{acta,custers}.  That the AF state of YbRh$_2$Si$_2$
lies on the FL side of the TQCP is confirmed experimentally by
verification that its thermodynamic and kinetic properties, such as
the specific heat and resistivity, behave in accordance with
the predictions of FL theory.

The proximity of a Fermi system to its TQCP creates the opportunity
for propagation of a transverse zero sound mode, henceforth referred to
as a {\it zeron} for brevity, which is known to propagate in 3D liquid
$^3$He.  Applicability of the zeron scenario to elucidation of properties
of 2D liquid $^3$He has been explored in Refs.~\cite{sound1,mig100,sound2}.
In general, the zeron mode provides an additional contribution to
the thermal conductivity $\kappa$ analogous to that coming from
phonons and/or magnons.  However, in contrast to the magnon mode,
the zeron contribution to $\kappa$ is quite insensitive to the
imposition of external magnetic fields $B$, a feature consistent with
experimental observations of the behavior of $\kappa (B)$ in
YbRh$_2$Si$_2$ \cite{arxiv,brison}.

A different mechanism for violation of the WF law comes into play on the
NFL side of the TQCP.  Beyond this point it is no longer the case that
variation of the ground-state energy $E$, given by the FL expression
\beq
\delta E=\sum \epsilon({\bf p})\delta n({\bf p})  ,
\label{nsc}
\eeq
yields a positive result for any admissible variation of the Landau
quasiparticle momentum distribution $n({\bf p})$.  This breakdown
triggers a rearrangement of the Landau state \cite{physrep}.  The
WF law can be profoundly affected if the rearrangement involves
a specific topological phase transition traditionally known as fermion
condensation, which gives rise to the formation of flat bands
\cite{ks,vol,noz,lee,kats}.  Emergence of a flat band entails
renormalization of conventional FL formulas, notably those in which
the residual electrical and thermal resistivities, respectively $\rho_0$
and $w_0$, are necessarily attributed to impurities.  It will be shown
that the impurity-independent flat-band contribution to $\rho_0$ is
always less than that to $w_0$.  Consequently, the net effect associated
with the flat-band scenario on the WF law is a {\it decline} of the ratio
$L(0)/L_0$ from unity, in agreement with available experimental
data \cite{brison,flouq,tanatar,dong}.

\paragraph{Violation of WF law: zeron scenario.}
We now turn to a more detailed analysis of the posited scenarios
for violation of the WF law, beginning with the zeron mechanism
for the recently observed downturn of the thermal resistivity.
On the FL side of the TQCP, the intervention of the zeron mode
can be addressed with the aid of the standard FL kinetic equation
\cite{lan1,lan2}
\beq
\left(\omega_k{-}{\bf k}\frac{\partial\epsilon({\bf p})}
{\partial {\bf p}} \right)\! \phi({\bf p},{\bf k})\!=\!
-{\bf k}\frac{\partial n({\bf p})}{\partial {\bf p}}\!\!\int\!\!\!
f({\bf p},{\bf p}_1)    \phi({\bf p}_1,{\bf k})d\upsilon_1,
\label{coll}
\eeq
in which $\phi({\bf p},{\bf k})$ describes the deviation of the quasiparticle
momentum distribution from equilibrium, $f({\bf p},{\bf p}_1)$ is
the Landau interaction function, and $d\upsilon$ is the volume element
in 3D momentum space, while $\omega_k=c_zk$. The impact of an external
field due to the crystal lattice will be discussed later.

By way of preparation, it is helpful first to consider the basic model of
a homogeneous Fermi liquid in which a single band crosses the Fermi surface.
In this case, the dispersion equation (\ref{coll}) for the zeron
group velocity $c_z$ is conveniently recast as \cite{halat}
\beq
1-\frac{6}{F_1}= 3\left(\frac{c^2_z}{v^2_F}-1\right)\left( \frac{c_z}{2v_F}
\ln\frac{c_z+v_F}{c_z-v_F}-1\right)
\label{de}
\eeq
in terms of the dimensionless first harmonic $F_1=f_1N(0)$ of the Landau
interaction function $f$, where $N(0)=p_FM^*/\pi^2$ is the density of
states of 3D homogeneous matter.  Here we may locate the TQCP itself
at the point where $v_F=p_F/M^*=0$, which is determined by
the relation \cite{lan1}
\beq
1=\frac{1}{3}f_1N_0(0),
\label{loc}
\eeq
in terms of the density of states $N_0(0)=p_FM/\pi^2$ of noninteracting
quasiparticles.  As seen from Eq.~(\ref{de}), a nontrivial solution does
exist provided $F_1>6$, or equivalently $M^*/M>3$.

Calculations are greatly facilitated near the TQCP where $F_1\gg 1$;
indeed they essentially coincide with those executed for longitudinal
zero sound in a neutral Fermi liquid.  Simple algebra leads to
\cite{halat}
\beq
c_z= \frac{v^0_F}{\sqrt{5M^*/M} }\propto v_F\sqrt{F_1}  ,
\label{cz}
\eeq
where $v^0_F=p_F/M$.  Just as for the longitudinal mode in a neutral
Fermi liquid, the transverse mode in question feels no Landau damping,
since $c_z>v_F$.   Moreover, its dissipation at $T=0$ is also related
to the generation of incoherent electron particle-hole ($ph$) pairs.
This feature allows us to adopt the same procedure in evaluating
the damping $\gamma_z(T)$ of the zeron mode by the $e-ph$ pair 
mechanism, as applied to evaluation of the damping of longitudinal
zero sound in neutral Fermi systems \cite{halat}.  The result is
$\gamma_z(T)=1/(c_z \tau(T))$, where $\tau(T)\propto T^{-2}$
is a time characteristic of the damping of single-particle excitations.
Hence $\gamma_z(T\to 0)\propto T^2$, implying that the scattering
length $l$ goes like $T^{-2}$, while the thermal conductivity
$\kappa\propto C(T)\,l(T)\,v_F$ varies as $T^{-1}$ \cite{lanlif} to produce
the behavior $w(T)=w_0+w_2T^2$.

The single-band model developed above must be modified to deal
with the actual case of YbRh$_2$Si$_2$, in which several electron bands
simultaneously cross the Fermi surface.  Among them, those associated with light
carriers of Fermi momenta $v_l$ play the major role.  The Fermi surface
of YbRh$_2$Si$_2$ is anisotropic, with heavy carriers occupying its
smaller portion.  In the simplest case, assumed here, the velocities
of light carriers are subsumed by a single parameter $v_l\gg v_F$,
and Eq.~(\ref{de}) is replaced by
$$
1-\frac{6}{F_1}= 3\left(\frac{c^2_z}{v^2_F}-1\right)\left(\frac {c_z}{2v_F}
\ln\frac{c_z+v_F}{c_z-v_F}-1\right) $$
\beq
+\frac{3v_F}{v_l} \left(\frac{c^2_z}{v^2_l}-1\right)\left(\frac {c_z}{2v_l}
\ln\frac{c_z+v_l}{c_z-v_l}-1\right)  .
\label{de2}
\eeq
It is readily seen that the zeron group velocity remains almost
unchanged.  Indeed, upon neglecting the contribution from the light band
to the real part of the right side of Eq.~(\ref{de2}) in view of the small
value of the prefactor $v_F/v_l$, we arrive at the same equation as
before for the real part $c_R$ of the zeron group velocity $c_z=c_R+ic_I$.
Thus, $c_R$ does in fact satisfy Eq.~(\ref{cz}).  A new feature of
the modified model is the presence of Landau damping of the zeron mode,
which comes into play because near the topological QCP, the zeron
group velocity $c_R$ is larger than the Fermi velocity $v_F$ of the
heavy band, while remaining less than that of the light band.

The inequality $c_R<v_l$ lifts the ban on zeron emission/absorption
by electrons, since in this case the conservation law
$\epsilon({\bf p}+{\bf k})-\epsilon({\bf p})-\omega_k=0$ is met
for the light carriers, thereby opening a channel for damping of
the zeron mode.  Observe then that in the last term on the right
side of Eq.~(\ref{de2}) we have
$\ln\left((c_z+v_l)/(c_z-v_l)\right)\simeq \ln (-1)=i\pi$,
because $|c_z|\ll v_l$.  This term cancels out the imaginary part
of the first term to establish
\beq
c_I\propto  (v^0_F/v_l)^2 \frac{p_F}{M^*} .
\eeq
Since $c_I\propto 1/M^*\ll c_R\propto \sqrt{1/M^*}$, the zeron mode
turns out to be weakly damped, much like the phonon mode in solids.

Pursuing this analogy, the kinetic theory of the electron-zeron problem
recapitulates that of electron-phonon physics. The only significant
difference naturally concerns the normalization of the electron-zeron
vertex part $g$.  This quantity is determined from analysis of the
equation for the amplitude $\Gamma(1,2;\omega)$ of electron-electron
scattering in the $ph$ channel near its pole $\omega=\omega_k$
\cite{migdal}.  In symbolic form one has
$\Gamma={\cal U }+\left({\cal U} GG\Gamma\right)$,
where ${\cal U }$ denotes the block of Feynman diagrams irreducible
in the $ph$ channel and $G$ is the single-particle Green function;
the brackets imply integration and summation over intermediate momentum
and spin variables.  This amplitude consists of a pole term
$\Gamma^P(1,2,\omega\to \omega_k)=g(1)g(2)/(\omega-\omega_k)$ and a
regular remainder $\Gamma^R(1,2,\omega)$ that satisfies the symbolic equation
$\Gamma^R={\cal U}+\left({\cal U}GG\Gamma^R\right)+\left({\cal U}X g\right)g$
where $X=\partial (GG)/\partial\omega$.  Upon multiplying both sides of
this equation by the product $gGG$ from the left and integration over
intermediate momenta and summation over spins, most of the terms cancel
out to yield $(g^2\partial (GG)/\partial \omega)=-1$ \cite{migdal}.
In the explicit form, it reads
\beq
g^2(k)\! \int\!\!  \frac{(n({\bf p}+{\bf k}){-}n({\bf p}) )
(\epsilon({\bf p}+{\bf k}){-}\epsilon({\bf p}) )}
{( \omega^2_k-(\epsilon({\bf p}+{\bf k}){-}\epsilon({\bf p}) )^2)^2}
\frac {2 d^3p}{(2\pi)^3} ={1\over 2\omega_k}.
\label{norm}
\eeq
Analysis of this relation is greatly simplified in the vicinity of the
topological QCP.  Since $c_z^2\gg v_F^2$, the difference
$\omega_k^2-(\epsilon({\bf p})+{\bf k})-(\epsilon({\bf p}))^2$
reduces to its first term, $ c_z^2 k^2$; after some algebra
we obtain
\beq
g^2(k) \simeq \frac{v^0_F v_F\omega_k}{p^3_F} .
\label{norm1}
\eeq

The zeron contribution $\kappa_z(T)$ to the conventional electron
thermal conductivity $\kappa_{ee}(T)\propto T^{-1}$ can be evaluated
with the aid of textbook formulas. In these, the phonon Debye temperature
is replaced by the zeron temperature $\Theta_z=c_zk_{\rm max}$, where
$k_{\rm max}$ denotes the maximum value of the zeron wave vector.
This yields (see e.g.\ \cite{lanlif})
\begin{eqnarray}
\kappa_z(T)&\propto& \Theta^2_z/T^2, \quad T<\Theta_z, \nonumber\\
\kappa_z(T)&=&{\rm const.}, \quad T>\Theta_z  .
\label{upt}
\end{eqnarray}
As $T\to 0$, the zeron part of $\kappa(T)$ is seen to soar upward more
rapidly than $\kappa_{ee}(T)$. Therefore $\Theta_z$ can in fact be
responsible for the upturn of the thermal conductivity $\kappa(T)$
at $T\to 0$ reported in Refs.~\cite{arxiv,brison}, {\it provided}
its value does not exceed $30\, {\rm mK}$.

The zeron-electron collision integral $ I_{ze}(T)$ is markedly enhanced
in the classical-like region $T\geq\Theta_z$ where all the zerons have
energies lower than $T$.  Setting $\epsilon'=\epsilon({\bf p}+{\bf k})$,
this integral takes the form
\cite{lanlif}
\beq
I_{ez}\propto \frac{\partial n_0(\epsilon)}{\partial \epsilon}
\!\!\int\!\! g^2(k)\frac{\partial N_z(\omega)}{\partial \omega}\delta
(\epsilon{-}\epsilon')\left(\varphi{-}\varphi'\right) \frac
{\omega d^3k}{(2\pi)^3}  ,
\label{colez}
\eeq
where $\varphi(\epsilon)$ and $\varphi'=\varphi(\epsilon')$ are the
deviations of the electron distribution function from its equilibrium
form $n_0(\epsilon) =(1+\exp(\epsilon/T))^{-1}$, and $g$ is again the
electron-zeron vertex part.  The equilibrium zeron distribution
function $N_z(\omega)$ is simply $T/\omega$.  Insertion of
Eq.~(\ref{norm1}) into Eq.~(\ref{colez}) and straightforward
algebraic steps lead to
\beq
I_{ez}(T\geq \Theta_z)\simeq \frac{k^2_{\rm max}}{p^2_F} \varphi .
\eeq
At $T\geq \Theta_z$, both the electron-zeron collision integral and
the zeron contribution to the thermal and electrical resistivities differ
from their electron-phonon counterparts \cite{lanlif} by the ratio
$k^2_{\rm max}/p^2_F$.

The magnitude of the factor $k^2_{\rm max}/p^2_F$ can be estimated by
tracing the evolution of the spectrum $\omega_z(k)$ as a function of
wave vector $k$. The key point is that the sign of the difference
$\omega_k^2-(\epsilon({\bf p}+{\bf k})-\epsilon({\bf p}))^2$ in
Eq.~(\ref{norm}) must be positive. Hence $\omega_k$ grows rapidly
as $\epsilon({\bf p}+{\bf k})-\epsilon({\bf p})$ increases, becoming
a nonlinear function of $k$ in the momentum region where the
spectrum $\epsilon({\bf p})$ ceases to be relatively flat. Such a
nonlinear behavior of $\omega_k$ triggers the decay of the zeron mode
at a critical wave vector $k_{cr}\simeq k_{\rm max}$ where the relation
$\omega(k_{cr})=\omega_{k_1}+\omega_{k_2}$, with $k_1+k_2=k_{cr}$,
is met for the first time, implying that the zeron mode becomes
unstable at $k>k_{cr}$.

Working within the zeron scenario, we now seek a resolution of
the puzzling low-$T$ behavior of the thermopower (or Seebeck
coefficient) of YbRh$_2$Si$_2$, as revealed in Ref.~\cite{stegt}.
According to conventional (i.e., FL) theory, the Seebeck
coefficient $S(T)$ must vary linearly with temperature
as $T \to 0$, i.e., $- S(T)=s_{FL}T$, with a positive
constant $s_{FL}$ \cite{barnard,lanlif,behnia}.  Instead, as
measured in this compound, the function $- S(T)$ exhibits a deep
downturn with a subsequent sign change that occurs {\it in the
same temperature region} where the abrupt upturn in thermal
conductivity was observed \cite{arxiv,brison}.

This coincidence suggests that the downturn is due to zeron drag,
in analogy with the well-known phonon drag.  Since input parameters
specifying the boson mode, such as the group velocity $c$ or the critical
wave vector $k_{\rm max}$, are absorbed into the corresponding Debye
temperature, the zeron contribution $S_z$ to the Seebeck $S(T)$ may
be written as \cite{barnard,lanlif}
\begin{eqnarray}
S_z\propto T^2/\Theta_z^2, \quad T<\Theta_z, \nonumber\\
S_z\propto \Theta_z/T, \quad T>\Theta_z .
\end{eqnarray}

In case the drag contribution to $S(T)$ and the standard FL contribution
have opposite signs, they may cancel each other, thereby
allowing the total Seebeck coefficient $S$ to change sign.  The
maximum $ S_z^{\rm max}$ of the zeron drag term, although small,
is nevertheless {\it temperature independent} \cite{lanlif}, while
the FL term $-s_{FL}T$ vanishes at $T\to 0$.  It then follows that
the total Seebeck coefficient $S$ becomes positive in some temperature
region, provided $\Theta_z$ meets the requirement
\beq
\Theta_z<  S_z^{\rm max}/s_{FL}  .
\label{ineqs}
\eeq
\begin{figure}[t]
\begin{center}
\includegraphics[width=0.3\textwidth,height=0.5
	\textwidth]{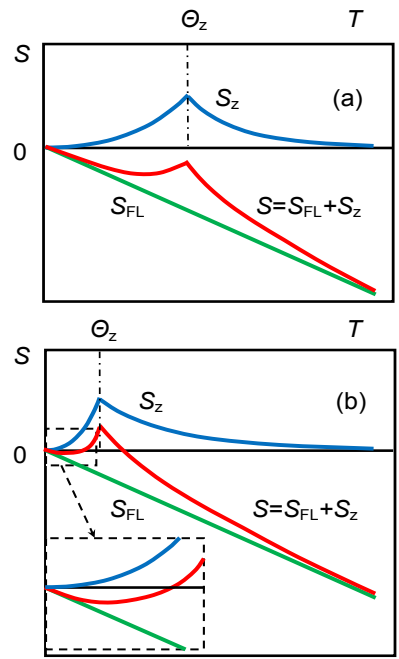}
\end{center}
\caption{ 
Schematic illustration of the total Seebeck coefficient $S$ (red line)
and its separate Fermi-Liquid $S_{FL}=-s_{FL}T$ (green line) and
zeron $S_z$ (blue lines) contributions.  In the case
$\Theta_z>S_z^{\rm max}/s_{FL}$ (panel (a)) the total
$S(T)=-s_{FL}T+S_z(T)$ does not change sign, while in
the opposite case $\Theta_z<S_z^{\rm max}/s_{FL}$ (panel (b))
it changes sign {\it twice}.
 }
\label{fig:Seebeck}
\end{figure}
Since the FL term in $S$ prevails at $T\to 0$, this implies that
at extremely low $\Theta_z$, the Seebeck coefficient $ S(T)$ must
change sign {\it twice}, (see Fig.~1).  This behavior provides
a fingerprint of the zeron scenario for the NFL component of the
thermopower.

The role of both these mechanisms becomes more profound in the presence
of a periodic external field of the crystal lattice, with Bloch-Wannier
wave functions as eigenfunctions of the solid-state Hamiltonian. As an
illustration, consider the well-known tight-binding model, in which the
Coulomb interaction between moving electrons and localized atoms is taken
into account, while interactions between electrons themselves are
neglected.  The single-particle spectrum of the simplest version of
this model, in which only nearest-neighbor matrix elements of the
electron-atom interaction Hamiltonian are included, has the form
$\epsilon({\bf p})\propto \cos p_xa+\cos p_ya+\cos p_za$, where $a$ is
the lattice constant. This formula implies that the inverse tight-binding
group velocity $1/|\partial\epsilon({\bf p})/\partial {\bf p}|$, which
enters the density of states $N_a(0)$ of noninteracting electrons
moving in the external field of the crystal lattice, diverges near
the saddle points $(0,\pi) \ldots  (\pi,0)$ as
$1/|\sin p_x^{vH}a|\simeq 1/| \sin p_y^{vH}a|\simeq 1/|\sin p_z^{vH}a|$,
where ${\bf p}^{vH}=(p_x^{vH},p_y^{vH},p_z^{vH}$) stands for the
coordinates of the van Hove point.  Since the compounds of interest
here have open Fermi surfaces crossing the Brillouin zone close to the
saddle points, such an enhancement is often substantial.

In dealing with strongly correlated electron systems, the product
$f_1N_a(0)$ must then be of order of unity (cf.~Eq.~(\ref{loc})),
assuming $f_1$ is again identified with the corresponding matrix element
of the effective $e$--$e$ interaction.   Consequently, when estimating
the suppression of the group velocity $c_z$, one can employ a modified
version
\beq
c_z= \frac{v^0_F}{\sqrt{5M^*M_a} }
\eeq
of Eq.~(\ref{cz}), in which the parameter $M_a$ specifies the density
of states $ N_a(0)$ in the same way as $M$ specifies the density of
states $N_0(0)$.

The enhancement of the density of states $N(0)$ of YbRh$_2$Si$_2$
beyond what is typical of conventional metals can be estimated
from experimental data on the specific-heat coefficient
$\gamma=C(T)/T=1.5\, {\rm J/ mole}\, {K}^2$.  In conventional
metals, the $e$--$e$ interaction is rather small, and the densities
of states of these systems, varying in the interval $1\div 5\,
{\rm mJ/mole}\, K^2$, are well approximated by the FL formula
$N_0(0)\propto M$.  We thus infer that the total enhancement factor
$N(0)/N_0(0)=M^*/M$ for YbRh$_2$Si$_2$ is about $10^3$. Presumably
the difference between $M^*$ and $M^*_a$ is not as large as that
between $M^*$ and $M$. If so, the ratio $c_z/v^0_F$ is suppressed
by a factor $\simeq 10^3$.

The impact of anisotropy effects on the magnitude of the ratio
$k^2_{\rm max}/p^2_F$ is likewise well pronounced, since
YbRh$_2$Si$_2$ has a typical open Fermi surface, with flattening
of the single-particle spectrum occurring solely in narrow regions
close to the van Hove points. This situation provides for an
additional decline of the characteristic zeron temperature
$\Theta_z$, down to values compatible with the $30\, {\rm mK}$
needed to explain the upturn of the thermal conductivity of
this compound \cite{arxiv,brison}.

Judging from the preceding theoretical development, analysis,
and results, the zeron scenario for explanation of the experimental
findings of Refs.~\cite{arxiv,brison} seems to be rather plausible.
If this scenario is indeed applicable, the fate of the WF law
in YbRh$_2$Si$_2$ at the lowest accessible temperatures can
be clarified only by subtracting the zeron contribution to
the thermal resistivity, as treated above.

It is noteworthy that in the vicinity of the TQCP where the
density of states $N(0)$ diverges, the number of collective modes
associated with Eq.~(\ref{coll}) and specified by two quantum numbers
(orbital momentum $l$ and its projection $m$), becomes rather
large. Accordingly, the task of finding the spectrum of these modes
becomes quite complicated near the topological QCP, where
many harmonics of the interaction function contribute to the
dispersion equation (\ref{coll}).  We address here the simple
case $m=l$ in homogeneous matter where the dispersion equation
reduces to \cite{halat}
\beq
\frac{F_l}{(2l)!}\int\limits_0^1 (P^l_l(x))^2\frac{x^2}{s^2-x^2}  dx=1  ,
\eeq
with $P^l_l(x)$ a spherical function and $F_l=f_lN(0)$.
For a solution $s=c/v_F>1$ to exist, the harmonic $f_l$ must be
positive. However, as seen from the case $l=1$ addressed above,
although this requirement is necessary, it is not sufficient (recall
that $s_{\rm min}=1$). In the present case, replacement of $x^2/(1-x^2)$
by $-1+1/(1-x^2)$ allows both the integrals involved to be performed
analytically, yielding
\beq
(F_l)^{cr}=(f_l)^{cr}p_FM^*/\pi^2=2l(2l+1) .
\label{ineq1}
\eeq
This result, setting the lower limit on the dimensionless parameter $F_l$
allowing propagation of the corresponding zero-sound mode, is in agreement
with $(F_1)^{cr}=6$ in the zeron case $l=1$.  As seen, the greater
the orbital momentum $l$, the larger the effective mass $M^*$ should
be to meet the condition involved. In this situation, the immediate
vicinity of the TQCP tends to be overcrowded by different zero
sounds, opening new boson channels that contribute weightily to
the thermal conductivity $\kappa$.

\paragraph{Violation of WF law: flat-band/Peierls scenario.}
Returning to the case of YbRh$_2$Si$_2$, we observe that upon subtracting
the posited zeron contribution to the thermal conductivity, the WF law
breaks down on the {\it disordered side of the AF transition} at
$T_N\leq T<1\, K$, the ratio $ L(T)/L_0$ {\it becoming less than unity}
\cite{arxiv,brison}.  It is presumably significant that other instances
of violation of the WF law documented in the literature \cite{tanatar,dong}
also involve a decline of the WF ratio.

This empirical conclusion is indicative of the failure of the standard
FL scenario, in which nonzero values of $\rho_0$ and $w_0$ are attributed
entirely to scattering of light carriers by impurities.  Its inadequacy
was revealed in measurements \cite{sidorov} of the electrical resistivity
of the normal state of the superconducting heavy-fermion metal CeCoIn$_5$
under pressure. These measurements showed a huge variation of the
residual resistivity $\rho_0(P)$ in a narrow region of the $T-P$ phase
diagram of this system near a critical pressure $P^*=1.6\, {\rm GPa}$,
even though the impurity number remained unchanged at any applied
pressure.

In Ref.~\cite{PRBcecoin5} such challenging behavior of CeCoIn$_5$
was attributed to fermion condensation (FC).  As indicated earlier, such
a state arises from a rearrangement of the Landau state at a TQCP
involving the formation of a flat band.  A flat band is composed of
the totality single-particle states that belong to a portion of
the spectrum that is completely flat at $T=0$, i.e., $\epsilon({\bf p})=0$,
and exhibits linear-in-$T$ dispersion at finite $T$ \cite{ks,vol,noz,lee,kats}
(for recent reviews, see \cite{an2012,book,volovik,yaf}). Since
quasiparticles in this state share the same energy, one has a
situation analogous that of Bose condensation; hence the original
name fermion condensate (FC) for the flat band.  A remarkable feature
of the FC/flat-band phenomenon is an inherent dramatic enhancement
of the density of states, with a corresponding amplification of the
superconducting critical temperature $T_c$ in systems hosting flat
bands.  This mechanism may in fact be relevant to the recent discovery
of high-$T_c$ superconductivity in sulfur hydride \cite{eremets}.

Flat bands do provide a natural explanation of the striking behavior
seen on the disordered side of the AF phase transition in the
nonsuperfluid compound YbRh$_2$Si$_2$ \cite{kss,kz,kzc}, where the
resistivity $\rho(T)$ shows linear-in-$T$ variation over an enormous
temperature range from $T_N=70\,{\rm mK}$ up to at least $7\,K$.  It
must be noted that all the numerous versions of the standard
critical-fluctuation Hertz-Millis approach fail to explain this
pivotal observation.

With regard to the fate of the WF law in electron systems of solids
possessing one or more flat bands, there exists a universal mechanism
for its violation that is absent on the FL side of the underlying
topological phase transition.  This mechanism is associated with
additional contributions to the conventional FL impurity-induced
residual resistivities $\rho_0$ and $w_0$ that result from elastic
scattering of light carriers by the heavy quasiparticles of flat
bands, in a process analogous to elastic scattering of light carriers
by impurities.  In both cases, contributions to the damping $\gamma(T)$
of single-particle excitations are $T$-independent and proportional
to the density of heavy carriers \cite{PRBcecoin5, shag1}.

Patently, if the two types of scattering were identical, the WF ratio
in systems possessing flat bands would keep the same value as in FL theory.
However, this is not the case, because the FC quasiparticles belong to
the same electron system as the normal ones.  It follows that in the
case of a homogeneous charged Fermi liquid having flat bands but
containing no impurities, for which the total electron momentum is
necessarily conserved, the solution of the kinetic equation acquires
a Peierls term $\delta n({\bf p})\propto {\bf p}$. This implies the
presence of an analogous term proportional to momentum ${\bf p}$
in the electrical current ${\bf J}$ as well; accordingly, the
electrical current flows without resistivity. As a result, in the
actual case where the impurity-induced quantities $\rho_0$ and
$w_0$ have nonzero values, $\rho_0$ remains unchanged through the
fermion-condensation phase transition, while the thermal resistivity
$w_0$, evaluated at ${\bf J}=0$, receives an addition contribution.
This contribution, associated with the enhancement of the damping
$\gamma(T)$, is proportional to the FC density. Thus, the WF ratio
should be {\it reduced from unity} in charged Fermi liquids hosting
flat bands.

The same situation prevails for electron systems of solids having
flat bands. This is true despite the fact that the electron momentum
is no longer conserved due to Umklapp processes, which provide for
momentum transfer to the crystal lattice and lead to a nonzero residual
resistivity $\rho_0$ even in the absence of impurities.  The magnitude
of $\rho_0$ then turns out to be proportional to the square of the
Umklapp integral for $e$--$e$ collisions \cite{lanlif}. (Other
processes, notably those involving collective modes, are ineffective
in the limit $T\to 0$ relevant to the WF ratio.)  The magnitude of
the resulting contribution to $\rho_0$ turns out to be rather small,
confirming the validity of the above prediction of a {\it decline} of the
WF ratio in systems with the flat bands.

Let us now consider, within the flat-band scenario, how the WF ratio
evolves under decrease of the impurity concentration.  With improvement
of purification efficiency, the influence of flat-band contributions
to $\rho_0$ and $w_0$ continues to grow.  Thus, starting at the critical
level at which flat-band contributions become comparable to those from
impurities, one expects to see a rapid enhancement of the decline
of the WF ratio with further reduction of impurity concentration.

It is illuminating to apply this flat-band/Peierls scenario to assess
the status of the WF law in electron systems subject to an external
magnetic field ${\bf B}$. The essential point is that the Peierls
contribution survives only through components of the current vector
${\bf J}$ parallel to the direction of the magnetic field ${\bf B}$.
One has only to observe that this contribution is nullified if ${\bf J}$
is perpendicular to ${\bf B}$, since in this case the operators
${\bf J}$ and ${\bf B}$ do not commute with each other.  Thus, even
if the WF law is violated at ${\bf B=0}$, imposition of a magnetic
field, perpendicular to ${\bf J}$ leads to its recovery, in agreement with experiment
\cite{paglione,flouq,tanatar,dong}.

\paragraph{Conclusions.}
We have proposed and explored two different scenarios, in some
sense complementary, for violation of the WF law in strongly correlated
electron systems of solids. The implicated regions of the Lifshitz
phase diagram lie in the neighborhood and on both sides of the TQCP at
which the density of states diverges and the conventional Landau state
loses its stability.
\begin{itemize}
\item[(i)]
{\it Fermi-Liquid side of the topological quantum critical point (TQCP): zeron scenario}.  We have
demonstrated that there emerges an additional channel of the thermal
conductivity that is mediated by a new branch of collective
excitations, a transverse zero-sound mode termed the zeron.
This mechanism gives rise to an enhancement of the WF ratio
$L(T)/L_0$ at finite temperatures on this side of the TQCP , while  as  $T$ goes down to 0, so that  $T<\Theta_z$, the WF ratio tends to 1, and the WF law is recovered.
\item[(ii)]
{\it Non-Fermi-Liquid side of TQCP: flat-band/Peierls scenario}.  A quite
different mechanism comes into play for those systems in which the
topological transition leads to the formation of one or more flat
bands featuring heavy quasiparticles. In this case an additional
channel emerges in the electrical conductivity, associated with the
occurrence of a Peierls term in the electrical current.  This
mechanism leads to a suppression  of the Lorenz number $L(T)$  at sufficiently low temperatures,  giving rise to  the decline of the WF ratio $L(T\to 0)/L_0$. In the classical temperature region  $T>\Theta_D$, this decline is eliminated,  because elastic  phonon contributions to both $\rho$ and $w$ prevail, so  that the WF law  is recovered.
\end{itemize}
The available experimental data \cite{arxiv,brison,paglione,tanatar,dong} are
consistent with the phenomenological analysis presented here.

\paragraph{Acknowledgements.}
The authors are grateful to J.\ Paglione and G.\ Volovik for valuable
discussions.  This research was partially supported by RFBR grants
13-02-00085 and 15-02-06261, and by grant NS-932.2014.2 from the
Russian Ministry of Sciences. VRS was supported by the Russian Science 
Foundation, Grant No.~14-22-00281. VAK thanks the McDonnell Center for the
Space Sciences for timely support. JWC is indebted to the University
of Madeira and its Centro de Ci\^encias Matem\'aticas for gracious
hospitality during a sabbatical residency.

\end{document}